\documentclass[usenatbib]{mnras}

\usepackage{apjfonts}
\usepackage{epsfig}
\usepackage{epstopdf}
\usepackage{graphicx}
\usepackage{natbib}
\usepackage{multirow}
\usepackage{array}
\usepackage{url}
\usepackage[T1]{fontenc}
\usepackage{aecompl}
\usepackage{placeins}

\title[Detection of Dispersed Radio Pulses]{Detection of Dispersed Radio Pulses: A machine learning approach to candidate identification and classification}

\author[T. Devine et al.]{Thomas Ryan Devine$^{1}$\thanks{E-mail: tdevine4@mix.wvu.edu}, Katerina Goseva-Popstojanova$^{1}$\thanks{E-mail: Katerina.Goseva@mail.wvu.edu} and Maura McLaughlin$^{2}$\thanks{E-mail: Maura.McLaughlin@mail.wvu.edu}\\
$^1$Lane Department of Computer Science and Electrical Engineering, West Virginia University\\
$^2$Department of Physics and Astronomy, West Virginia University}
\begin{document}
\maketitle
\begin{abstract}

Searching for extraterrestrial, transient signals in astronomical data sets is an active area of current research. However, machine learning techniques are lacking in the literature concerning single-pulse detection. This paper presents a new, two-stage approach for identifying and classifying dispersed pulse groups (DPGs) in single-pulse search output. The first stage identified DPGs and extracted features to characterize them using a new peak identification algorithm which tracks sloping tendencies around local maxima in plots of signal-to-noise ratio vs. dispersion measure. The second stage used supervised machine learning to classify DPGs. We created four benchmark data sets: one unbalanced and three balanced versions using three different imbalance treatments. We empirically evaluated 48 classifiers by training and testing binary and multiclass versions of six machine learning algorithms on each of the four benchmark versions. While each classifier had advantages and disadvantages, all classifiers with imbalance treatments had higher recall values than those with unbalanced data, regardless of the machine learning algorithm used. Based on the benchmarking results, we selected a subset of classifiers to classify the full, unlabelled data set of over 1.5 million DPGs identified in 42,405 observations made by the Green Bank Telescope. Overall, the classifiers using a multiclass ensemble tree learner in combination with two oversampling imbalance treatments were the most efficient; they identified additional known pulsars not in the benchmark data set and provided six potential discoveries, with significantly less false positives than the other classifiers. 

\end{abstract}

\section{Introduction}
\label{sect:intro}
This work focuses on the identification and classification of extraterrestrial, transient radio signals. In particular, we are concerned with transient, dispersed, radio signals as expected from pulsars, rotating radio transients (RRATs), and isolated events such as the quickly growing group of fast radio bursts (FRBs). Pulsars are rapidly spinning neutron stars which emit radiation from their magnetic poles \citep{handbook}. If those emission beams sweep past the Earth, they can be detected as ``pulses'' of radiation with extremely regular periods. RRATs, first discovered by \cite{2006Natur.439..817M}, are thought to be a special type of sporadically emitting pulsar. Throughout this paper, we will use the term `pulsar' to describe pulsars and RRATs. FRBs are bright, isolated radio bursts with millisecond durations and likely have extragalactic origins \citep{Lorimer02112007}. The study of pulsars and FRBs provides information about the extreme physics of neutron stars, and their unique properties allow a range of scientific applications.

There are two main approaches to pulsar detection in radio data: periodicity searches and single-pulse searches. Both techniques operate by searching the dedispersed raw data from radio telescope receivers, and produce output in the form of plots of candidate signals. The two searches differ in the types of phenomena they attempt to detect. Periodicity searches transform the time series into the frequency domain by applying Fast Fourier Transforms (FFTs) to identify periodic signals. The original time series data are then folded \citep{1996A&AS..117..197L} at the identified periods to amplify the signal-to-noise ratios (SNRs) of the periodic signals. Single-pulse searches, on the other hand, do not use FFTs or fold the data. This has the advantage of being able to detect strong, non-periodic signals that periodicity searches cannot detect. RRATs, for instance, are only detectable through their isolated pulses. However, single-pulse searches typically are not able to detect very regular, weak signals that would show up in a periodicity search.

Pulsar discoveries have been made through a variety of detection techniques. Despite all of these discoveries, \cite{2006ApJ...643..332F} theorized that the over 2,500 known pulsars comprise a small percentage of the potentially detectable pulsars in our galaxy. Furthermore, \cite{2011MNRAS.418..477B} projected that additional pulsars may be detectable in globular clusters. Discovering these pulsars, however, is very challenging. The signals are faint, requiring sensitive observations. Searches must deal with issues such as noise (resulting from receivers and the sky), radio frequency interference (RFI), and imbalanced data sets (i.e., only a very small fraction of the radio signals received originate from pulsars).

Traditionally, pulsars are discovered by manual inspection of the candidates produced by periodicity or single-pulse searches. Manual inspection by domain experts, to some extent, will likely always be integral to the pulsar discovery process. However, automation of the majority of the process is vital for the future of radio astronomy. Next generation instruments, such as South Africa's Karoo Array Telescope (MeerKAT) \citep{ska}, which is a precursor of the Square Kilometre Array (SKA), or the Five-hundred meter Aperture Spherical Telescope (FAST) \citep{fast} in China will have many more beams than the current instruments, resulting in significantly larger data sets. Automated approaches are the only feasible way to deal with big data, and offer many potential advantages to streamline the discovery process, e.g., by triggering the rapid follow up of candidates at multiple wavelengths to constrain their origins. Machine learning\footnote{These three terms are often used interchangeably.} algorithms have been applied to automatically detect pulsars in periodicity searches \citep{2010MNRAS.407.2443E,2012MNRAS.427.1052B,2014ApJ...781..117Z,2014MNRAS.443.1651M}. However, machine learning approaches with single-pulse candidates are lacking.

This paper presents a novel, two-stage approach to semi-automatic discovery of transient radio signals within the candidates produced by single-pulse searches. These transient signals are in the form of dispersed pulse groups (DPGs), which are collections of pulses appearing as peaks in the signal-to-noise ratio (SNR) vs dispersion measure (DM) subplot of a candidate plot, such as the one shown in Figure~\ref{fig:plot1}. Note that a DPG is different from a candidate in the traditional sense. A single candidate plot could potentially have many identified DPGs, since a DPG is any local peak in the SNR vs DM subplot. For the first stage, DPG identification, we present a new Recursive Algorithm for Peak IDentification (RAPID) which effectively identified pulsar signals. Individual DPGs, along with their characteristic features, served as instances for machine learning. For the second stage, DPG classification, we created binary and multiclass machine learning models to classify DPGs as pulsars or non-pulsars. The data used for this work were derived from the 350-MHz drift-scan survey performed with the GBT from May through August in 2011. The survey was conducted while the GBT was immobilised for refurbishing. The receivers remained active throughout the repairs and collected data at a radio frequency of 350 MHz as the sky passed through the beam of the telescope \citep{2013ApJ...763...80B}.


The remainder of this paper begins by giving a general background on pulsar searching in Section~\ref{sect:background}. Section~\ref{sect:related} provides the related work on pulsar searching and peak identification. We describe our two-stage approach in detail in Section~\ref{sect:ml}, and provide the results of our experiments in Section~\ref{sect:results}. Finally, Section~\ref{sect:conc} presents the conclusion.
\section{Background on Pulsar Searching}
\label{sect:background}

Pulsar discovery in radio data sets is typically approached in four phases: collection, dedispersion, periodicity or single-pulse search, and manual inspection. In the first phase, raw data are collected at radio telescopes as a time-series of voltages. A thorough description of the second phase, dedispersion, is given by \citet{handbook}, and will only be described briefly here. As a pulsar's radiation propagates through the interstellar medium (ISM), the ISM causes the pulses to be dispersed, with lower frequency components of pulses arriving later than higher frequency components. The time delay between two frequencies depends on three things: the difference between the frequencies of the observations, the observational frequency, and the DM, which is the integrated number of free electrons along the line of sight measured in $\rm pc~cm^{-3}$. Dedispersion is the process of removing these frequency-dependent delays.

\begin{figure}
	\centering
	\includegraphics[width=1.0\linewidth]{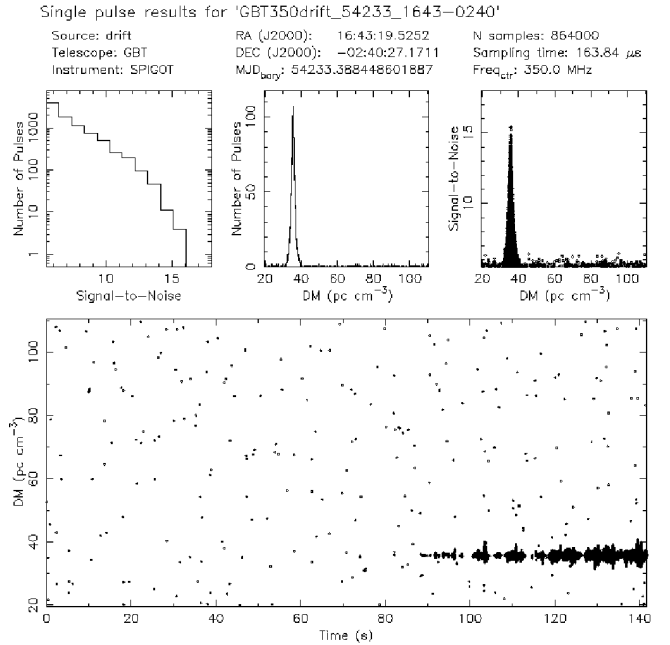}
	\caption{The known pulsar J1645--0317 identified by a \textit{single-pulse} search. The subplots, created using the software Presto \citep{2001PhDT.......123R}, clockwise from the top left, show a histogram of the number of pulses detected at different SNR ranges, a histogram of the number of pulses detected at each DM, a scatter diagram showing the SNR and DM of all pulses, and a scatter diagram showing the DM and time of each detected pulse, with the markers for individual pulses scaled in size by their SNR.}
	\label{fig:plot1}
\end{figure}

In the third phase, either periodicity or single-pulse searches are performed at a number of trial DM values, as discussed in Section~\ref{sect:intro}. The fourth and final phase traditionally consisted of manually inspecting a number of candidate plots, created by the software Presto \citep{2001PhDT.......123R}. Figure~\ref{fig:plot1} offers a single-pulse search candidate plot of a known pulsar. This type of plot contains four subplots: the top left is a histogram of the number of pulses for each SNR value, the top middle is a histogram of the number of observed signals for each DM, the top right is a scatter diagram showing the SNR values of any recorded pulses for each DM, and the bottom is a scatter diagram which shows the DM on the y-axis and the time each signal was recorded on the x-axis. In the bottom subplot, each point's size is scaled by the magnitude of its SNR value, i.e., larger SNR values appear as larger points \citep{0004-637X-596-2-1142}.

\section{Related Work}
\label{sect:related}

This section presents a brief survey of the literature pertaining to our work. The first subsection presents pulsar searches within single-pulse search output, the second reviews classification studies performed using periodicity searches, and the third contrasts several existing peak identification algorithms with our proposed technique, RAPID.

\subsection{Related Work on Single-pulse Searches}

\cite{0004-637X-596-2-1142} first presented a theoretical framework for performing single-pulse searches to detect fast radio transient signals. Their proposed automated detection approach first removed the baseline from the dedispersed data, then utilized an N-sample boxcar filter to detect significant events, which they plotted for manual inspection. They also explored a range of widths by adding a number of adjacent samples, smoothing the data, and then searching for single bright pulses. To avoid bias towards very strong individual pulses, they removed the brightest pulses in the first pass, then searched again. They recorded the DM, the arrival time of the pulse relative to the start of the observation, and the SNR for each pulse subsequently detected. The methodology presented by this work was adopted, in one form or another, in each of the following papers.

\cite{0004-637X-703-2-2259} presented results from radio transient searches using data from the seven beam Pulsar Arecibo L-band Feed Array (PALFA) survey \citep{0004-637X-637-1-446}. They performed matched filtering similar to \cite{0004-637X-596-2-1142}, but with a more sophisticated RFI excision scheme. Their search was customized to remove two types of RFI: RFI from radar unique to Arecibo (from the San Juan airport), and RFI simultaneously detected in several beams. They also used a friends-of-friends clustering algorithm, which formed the initial clusters by searching for events above a given SNR threshold, then added to the clusters by including adjacent samples above a given threshold. The brightest sample of a cluster was recorded as the cluster amplitude and the number of samples as the width. This search was less sensitive to weak, narrow pulses but resulted in a significant reduction of RFI events and resulted in seven pulsar discoveries.

\cite{2010MNRAS.401.1057K} performed a re-analysis of the Parkes Multi-beam Pulsar Survey (PMPS)\footnote{The PMPS \citep{2001MNRAS.328...17M} was completed between 1997 and 2003 using the Parkes radio telescope in Australia and is the most successful large-scale pulsar survey to date.} and discovered ten RRATs, suggesting that the population of transient radio-emitting neutron stars may be larger than initial predictions. They searched for bright single-pulses using matched filtering, as in \cite{0004-637X-596-2-1142}. To eliminate RFI, they used the ``zero-DM filter'', developed by \cite{MNR:MNR14524}, and also removed multi-beam events from consideration, as in \cite{0004-637X-703-2-2259}. They produced diagnostic plots for manual inspection and classification.

\cite{2011MNRAS.416.2465B} presented the initial results for an examination of the High Time Resolution Universe (HTRU) survey using similar search techniques. They stored parameter values in a database, which was then queried to see if events have more than two members and peak at a DM over 1.5 $\rm pc~cm^{-3}$. If so, summary plots were created of the events for manual assessment. Their efforts resulted in 11 discoveries of sparsely emitting neutron stars.

\cite{2012MNRAS.425.2501B} searched the archival PMPS data for RRATs, FRBs, and perytons (an unusual form of RFI detected in all 13 beams of the PMPS and other surveys). Their search followed the methodology of \cite{0004-637X-596-2-1142} and resulted in no detections of RRATs or FRBs, but did detect four
peryton-like events.

Using an iterative process to extract individual pulses, \cite{2013MNRAS.428.2857R} detected several single-pulse events, some of which were repetitive, in a search of the Andromeda Galaxy and its satellites with the Westerbork Synthesis Radio Telescope.

Most recently, \cite{chen} searched for RRATs in data from the GBT 350-MHz drift-scan survey. After applying similar filtering techniques, they grouped the data according to their relative positions in the DM vs time space and divided each group into five bins. The neighboring bins were then checked to see if the maximum SNR in each one was monotonically decreasing and created diagnostic plots for manual inspection. This work resulted in the discovery of 18 RRATs.

The papers presented above all include automated search techniques using heuristics, e.g. sifting candidates by known SNR or DM thresholds. Our work differs from the literature by not relying on heuristic thresholds to identify peaks, and by using supervised machine learning to develop intelligent classifiers.

\subsection{Related Work on Periodicity Searches}
\label{sect:rw-periodicity}

Classification techniques in the literature for periodicity search candidates include both automatic heuristic approaches\footnote{See \cite{Faulkner21112004,2009MNRAS.395..837K,2013MNRAS.433..688L}.} and machine learning approaches. As our focus is on machine learning, we only provide reviews of papers that use machine learning techniques \citep{2010MNRAS.407.2443E,2012MNRAS.427.1052B,2014ApJ...781..117Z,2014MNRAS.443.1651M}. The fact that these papers were all published in the last five years indicates that intelligent algorithms are becoming the new standard for pulsar classification.

\citet{2010MNRAS.407.2443E} used artificial neural networks (ANNs) to automate pulsar detection in the PMPS. They used a set of twelve features, including the pulse profile SNR, pulse profile width, and $\chi^2$ values of fits to theoretically optimal curves. Their training set consisted of 259 examples of known pulsars combined with 1,625 non-pulsar examples of noise or RFI. Their model led to the discovery of one pulsar.

\citet{2012MNRAS.427.1052B} also used ANNs to classify candidates. They expanded the input features from \citet{2009MNRAS.395..837K} and \citet{2010MNRAS.407.2443E} to include $\chi^2$ values for fits of the pulse profile to Gaussians and sinusoids, and profile histogram tests. Their resulting ANN was able to detect 85\% of pulsars in controlled tests with data from the HTRU survey. It was further found that the ANN's classifications depended on the training data used, leading them to recommend a representative sample of pulsars to increase the accuracy of the learner. This work resulted in the discovery of 75 pulsars.

Recently, \citet{2014ApJ...781..117Z} created an artificial intelligence program to identify pulsars using image recognition algorithms called the Pulsar Image-based Classification System (PICS). PICS consists of two layers and was designed to emulate a human expert's visual identification process. The first layer is a group of trained image learners (ANNs, convolutional neural networks (CNNs), and SVMs) which examine and score candidate subplots. These scores are combined using a logistic regression model to minimize classification errors in the training data. The PICS AI system was tested on the Green Bank North Celestial Cap pulsar survey and is currently integrated with the PALFA survey, where it has discovered six pulsars.

Another recent work by \citet{2014MNRAS.443.1651M} presented the classification results from a pulsar ranking system called Straight-forward Pulsar Identification using Neural Networks (SPINN). SPINN uses a customized ANN trained on 1,196 observations of pulsars from the HTRU all-sky pulsar survey combined with 90,000 randomly selected negative observations. They were able to correctly classify all known observations of pulsars in the HTRU data while reducing the number of candidates requiring manual inspection by several orders of magnitude. This system was responsible for the discovery of four pulsars.

\subsection{Related Work on Peak Identification}
\label{sect:rw_peaks}

Peak or trough identification is a common problem in many fields that require signal processing. Many different techniques have been proposed to solve this problem, ranging from general solutions to solutions highly specific to particular fields. In this section, we briefly discuss several existing peak identification approaches and describe why a new technique was required to identify DPGs in single-pulse search candidates.

A common approach for identifying peaks in time series data is to detect local maxima by noting sign changes in the slopes between a single point and its immediate neighbors. A major problem for this and all peak detection algorithms is their sensitivity to noise. Another popular solution is to first smooth the data with some sort of filter and then fit a given function to it \citep{citeulike:9398436}.

In mass spectroscopy, peaks have specific shapes. Taking advantage of this fact, \citet{Du01092006} developed a pattern matching algorithm using continuous wavelet transforms (CWTs). The basic shape of the peak was assigned to the wavelet function, which was in turn used to compute an array of CWT coefficients according to multiple scales. Peaks were then identified as ``ridges'' formed in the wavelet space.

\citet{4566694} proposed an algorithm to detect peaks and troughs based on momentum. The ``momentum'' was found by taking the product of the value of a data point and the rate of change at that point. A theoretical ball was then ``rolled'' from a known peak. As the ball descends the peak, its momentum increases and then decreases as it climbs another peak. When the momentum reaches zero, the ball was considered to have come to rest and that point is declared a new peak. Momentum changes are also affected by Newton-esque laws of motion, such as an analogue to friction.

In astrophysics, several burst detection algorithms have been proposed to identify gamma-ray bursts (GRBs). The Li-Fenimore algorithm (LFA) operated by binning the data and then labeling as a candidate peak each bin that had more counts than its immediate neighbors \citet{LFA}. A search was then conducted for each candidate peak to determine if the counts for non-immediate neighbors (more than one bin away) continued to diminish according to a given formula.

\citet{Zhu} proposed a burst detection algorithm to identify GRBs in real time. Their algorithm relied on wavelets by introducing a new data structure called the shifted wavelet tree (SWT), which was used to organize wavelet coefficients and additional information about the window by resolutions and time scales. The elastic window was created by automatically scanning different time resolutions and sizes and determining the window size accordingly.

\citet{Guidorzi201554} developed MEPSA, an algorithm similar to LFA that also used binning and the counts of signals in each bin to detect GRBs. MEPSA utilized 39 user-defined patterns to help peak identification. For each bin, the adjacent bins were searched to see if they fit any of the different patterns. MEPSA was more reliable than LFA, but came with an added overhead of 39 separate pattern comparisons for each possible peak.

We created RAPID because machine learning for DPGs has several requirements, and none of the algorithms listed above satisfied all of these requirements. First, identifying the peak alone is not sufficient; pertinent features must be extracted which include the shape and number of pulses in the range of the DPG, from beginning, to peak, to end. This necessitates that any detection routine be capable of identifying more than just local maxima in the data. Second, focusing on single values and ignoring local trends could overlook peaks in the data. Third, not all DPGs have a set width or height and are often very faint or buried in noise. Finally, due to the large quantity of radio data to be examined, the algorithm used for detecting DPGs must be simple and efficient, ideally making only one pass through the data.
\section{Our Proposed Machine Learning Approach}
\label{sect:ml}

\begin{figure}
	\centering
	\includegraphics[width=0.9\linewidth]{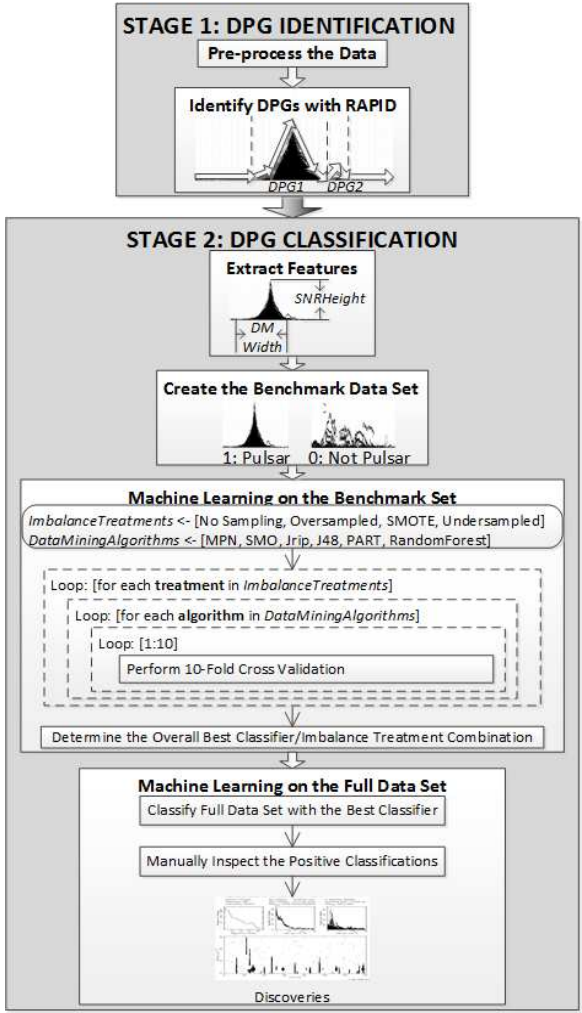}
	\caption{Our new machine learning approach to single-pulse detection.}
	\label{fig:approach}
\end{figure}

Our approach consists of DPG identification and DPG classification using machine learning with imbalance considerations.  Figure~\ref{fig:approach} provides an illustration of this process. Here, we detail the various stages.

\subsection{Data Pre-processing}
\label{sect:data}

Our initial data consisted of output from Presto's $single\_pulse\_search.py$ on data from the GBT drift-scan. The data were composed of individual files for 5,766 DMs (ranging from 0 -- 1,096 $\rm pc~cm^{-3}$) for 42,405 separate observation positions. Each file contained data describing the SNR, the downfact (a proxy for pulse width), and the time of each single-pulse event for that particular DM. These 244.5 million files encompassed 118 GB of data which required over 922 GB of storage space. Since single-pulse detections are often sparse, many of these files contained little or no data. However, each empty or sparse file still required four KB of storage space due to the minimum allocation size of most hard drives. From the 5,766 files for each observation position, we extracted only the data necessary for our research and created four large files, one for each data structure used by RAPID: number of pulses vs DM, SNR vs DM, maximum SNR vs DM, and DM vs time. This effectively eliminated data bloat by reducing the actual data size by a factor of 3.17 (from 118 GB to 37.2 GB), and the storage space required by a factor of 24.6 (from 922 GB to 37.5 GB). Additionally, runtime efficiency was improved by eliminating the need to parse text files and reducing the number of read operations to different locations on the hard drive.

\subsection{DPG Identification with RAPID}
\label{sect:repeat}

We propose a simple, recursive peak identification algorithm, RAPID, which divides its input into bins and performs linear regression \citep{agresti2009statistics} to fit a straight line to the points within each bin. The slopes of the fit lines for the bins are used to identify the larger slope trends of a DPG. In this context, a DPG is an instance of a peak and its surrounding decreasing values in the data used to create the integrated ``SNR vs DM'' subplot of a candidate plot, as in Figure~\ref{fig:plot1}. Note that a single candidate plot can contain many different DPGs, and only one of those DPGs will actually represent a pulsar. At this stage of our work, RAPID looks only at the maximum SNR values for each DM, not at individual pulses. 

RAPID can be tuned by adjusting two parameters, the \textit{bin size} and the \textit{slope threshold}. The bin size determines how smoothed the detected slopes will be. A smaller bin size allows the identification of narrower DPGs that could be missed by large bin sizes, but at the cost of increasing the size of the output and potentially missing wider DPGs. Using larger bin sizes smooths the data to ignore tiny fluctuations resulting from noise, but may miss smaller DPGs. The slope threshold is a limit placed on the rate of change between the maximum integrated SNR and the DM, and defines the minimum fit-line slope (FLS) required to consider a bin's trend as increasing or decreasing. Higher values will require steeper slopes for DPG recognition, and lower values will allow the detection of more gradual slopes. Strictly speaking, the SNR vs DM curve for a particular pulsar is dependent on the width, observing frequency, and distance from the central DM ($\delta \textrm{DM}$) of the pulse (as given by Equation~\ref{eq:fit}). However, at this point the width of the DPG is not known and we need an initial guess for the slope to begin our search. We set the slope-threshold at 0.5 so as not to exclude any gently sloping pulsars and still be able to identify those with steeper slopes. 

For each DPG, RAPID identifies: 1) the \textit{start}, the starting DM of the first bin to have a positive FLS greater than the slope threshold and immediately following two or more flat bins (bins with FLSs below the slope threshold) or one bin with a negative FLS, 2) the \textit{peak}, or maximum value between the start and the end, and 3) the \textit{end}, the starting DM of either the first single bin with a positive FLS or the first of two flat bins seen after the peak. Each bin FLS can take one of three values, depending on the slope threshold: \textit{1} -- positive and steeper than the slope threshold, \textit{0} -- shallower than the slope threshold, or \textit{-1} -- negative and steeper than the slope threshold. In this way, the algorithm determines if it is climbing or descending a DPG, if it has crossed the peak yet, or if it is on level ground. For example, if the preceding bin had an increasing slope, and the current bin's slope is decreasing, RAPID knows that it has climbed up to a peak and is now descending. If the next two bins were both below the slope threshold, then the algorithm would know that it had reached a termination point and would record the relevant data from the start to the end. By using sloping trends to find the starting and ending points of DPGs, RAPID can identify DPGs of various widths in only one pass through the data.

For each bin, the algorithm passes three values: \textit{starting DM} -- used to determine the next bin, \textit{current FLS} -- for comparison to the next bin's FLS, and \textit{status} -- keeps track of whether the signal has begun ascending and whether it has crossed a peak yet. The algorithm is recursive, in that it calls itself with each bin's calculation. This is more efficient in terms of memory and execution time when compared to a non-recursive implementation (using `while' loops) which ran approximately five times slower and used over eight times more memory.

RAPID is similar to the ``momentum'' peak identification technique proposed by \citet{4566694} and described in Section~\ref{sect:rw_peaks}. However, while their momentum technique relies on the instantaneous rate of change at a point, RAPID uses the slope of regression lines for bins of data points. By breaking the data into  bins, we eliminate the need for fitting a more complex equation, and calculations of its derivative, that could be  thrown off by noise or RFI. We also ensure that small fluctuations do not affect the overall trends.

RAPID also differs from other binning techniques for burst detection or peak identification \citep{LFA,Guidorzi201554,chen} in several key ways. First, all other binning techniques look at only a single value for each bin. If applied to DPG identification, one could use some value, say the mean, to represent the bin. However, this cannot tell us  which direction the points inside one bin are trending. Additionally, RAPID only needs to make one pass through the data, while LFA \citep{LFA} and MEPSA \citep{Guidorzi201554} perform an initial pass to identify candidates and then another pass to search for patterns.

Finally, RAPID is designed for a multi-threaded implementation to allow parallel execution. Since the data for each sky position are independent, RAPID can be instantiated in multiple threads to process the data from multiple positions simultaneously. The output from each scan for DPGs is saved individually and the results are aggregated when all scans are completed. 

\subsection{Feature Extraction}
\label{sect:features}

\begin{table*}
	\caption{Features extracted for each DPG and used by  machine learning algorithms for classification. Features 5 -- 8 were taken from data in the Pulse Counts vs DM plot in Figure~\ref{fig:plot1}, while the rest of the features were taken from the SNR vs DM plot in Figure~\ref{fig:plot1}.}
	\centering
	\footnotesize
		\begin{tabular}{||c|p{0.19\linewidth}|p{0.71\linewidth}||}
			\hline
			\hline
			\textbf & {Feature} & \textbf{Description} \\
			\hline
			1 & \textit{StartDM} & The starting DM of the DPG \\
			2 & \textit{StopDM} & The ending DM of the DPG \\
			3 & \textit{DMWidth} & \textit{StopDM} - \textit{StartDM}, or the width in DM of the DPG. \\
			4 & \textit{MaxPulseCount} & The maximum number of pulses occurring at a DM in the DPG. \\
			5 & \textit{IntegratedPulseCount} & The total number of pulses counted in the DPG. \\
			6 & \textit{AvgPulseCount} & The mean number of pulses detected per DM increment in the DPG. \\
			7 & \textit{PulseCountLocalPeakHeight} & \textit{MaxPulseCount} - \textit{AvgPulseCount}, or the height of the peak above the local average count of pulses in the DPG. \\
			8 & \textit{PulseCountPeakDM} & The DM corresponding to the maximum pulse count in the DPG. \\
			9 & \textit{MaxSNR} & The local maximum of the SNR values. \\
			10 & \textit{IntegratedSNR} & The sum of all SNRs recorded over the DPG. \\
			11 & \textit{AvgSNR} & The mean SNR value detected per DM increment in the DPG. \\
			12 & \textit{SNRLocalPeakHeight} & \textit{MaxSNRHeight}-\textit{AvgSNR}, or the height of the SNR peak above the local SNR average in the DPG. \\
			13 & \textit{SNRPeakDM} & The DM corresponding to the maximum SNR value in the DPG. \\
			14 & \textit{FittedMaxSNR} & The fitted value for $S$ in Equation~\ref{eq:fit}. \\
			15 & \textit{FittedWidth} & The fitted value for $w$ in Equation~\ref{eq:fit}. \\
			16 & \textit{$SNRMax\chi^2$} & The $\chi^2$ of the maximum SNRs recorded for the DPG against the ideal distribution, as per Equation~\ref{eq:fit}. \\
			\hline
			\hline
		\end{tabular}
	\label{tab:features}
\end{table*}

Once RAPID identifies a DPG, our code automatically extracts features to characterize it. The features are extracted from the data used to produce two subplots shown in Figure~\ref{fig:plot1}: the number of pulses (pulse counts) vs DM histogram and the SNR vs DM diagram, and are listed in Table~\ref{tab:features}. The features include measures of width and height, integrations to give an idea of the total ``strength'' of the DPG, and average values for the DPG.

The last three features in Table~\ref{tab:features} describe how well a DPG's shape in the SNR vs DM space fits the ideal theoretical shape of a single dispersed pulse \citep{0004-637X-596-2-1142}. Theoretically, the flux, which is proportional to the SNR, at some offset from the true DM, $\delta \textrm{DM}$, will follow Equation~(\ref{eq:fit}). Note that Equation~(\ref{eq:fit}) describes the shape of a single dispersed pulse, not a DPG. However, typically a group of dispersed pulses will be dominated by its brightest member, making a fit comparison to Equation~(\ref{eq:fit}) relevant.
\begin{equation}
  \frac{S(\delta \textrm{DM})}{S}=\frac{\sqrt{\pi}}{2} \zeta^{-1} \textrm{erf}\zeta
  \label{eq:fit}
\end{equation}
In Equation~\ref{eq:fit}, $S(\delta$ DM$)/S$ is the ratio of the observed flux to the peak flux, erf$\zeta$ is the error function, and $\zeta$ is the value given by:
\begin{equation}
  \zeta = 6.91 \times{10^{-3}} \delta \textrm{DM} \frac{\delta\nu}{w \nu^3},
  \label{eq:zeta}
\end{equation}
where $\delta\nu$ is the total bandwidth in MHz, $\nu^3$ is the cube of the central frequency in GHz, and $w$ is the full width in ms of the pulse at half of $S$ (FWHM).

\begin{figure}
	\centering
		\includegraphics[width=0.9\linewidth]{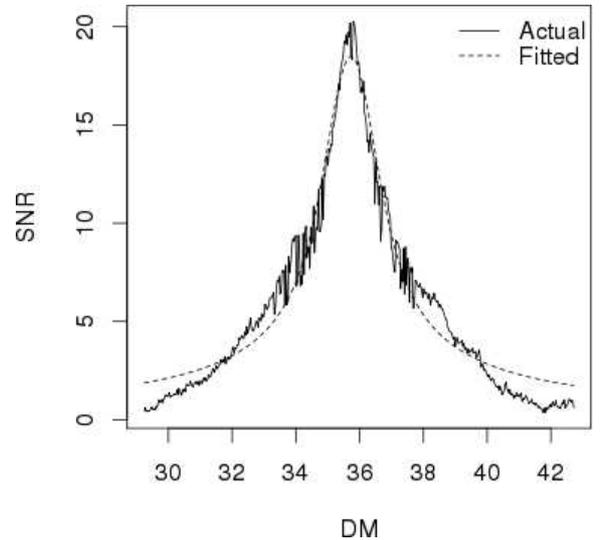}
\caption{The maximum SNR values (solid line) plotted against the calculated fit (dashed line) according to Equation~\ref{eq:fit} for the DPG representing the known pulsar J1645--0317.}
\label{fig:fit}
\end{figure}

We quantified how well each given distribution of points fits the theoretical shape by performing a non-linear least squares regression using Gauss-Newton optimization\footnote{We originally used a Levenberg-Marquardt optimizer, but it consistently required thousands of iterations to converge. The Gauss-Newton optimizer converged much more rapidly, drastically reducing the computation time.}, and required the difference between the root mean squared error of the current and previous iterations to be less than $10^{-4}$. We used the regression line to estimate $S$ and $w$ for each DPG and then compared the actual fitted curve to the expected theoretical curve by computing the $\chi^2$ value. Figure~\ref{fig:fit} provides an example plot of the fit line found for the DPG representing the known pulsar J1645--0317.

The features extracted for all DPGs identified by RAPID were saved in a data set referred to as the full data set throughout this paper.

\subsection{Creating the Benchmark Data Set}
\label{sect:labels}

In this paper, we use a supervised learning approach, which uses known positive instances (in our case, pulsars) to build a classification model. This requires training on a fully labelled data set where the class value of every instance is known \textit{a priori}. For validation, every instance must be manually inspected. The size of our full data set prohibited the labeling and use of all the instances. Instead, we identified as many DPGs representing known pulsars as possible, and combined them with a random sample of manually validated non-pulsar DPGs from the full data set.
To select the DPGs from our full data set that corresponded to known pulsars, we compared with the positions and DMs of the 2,234 pulsars listed in  the ATNF Pulsar Catalogue \citep{psrcat-paper}\footnote{http://www.atnf.csiro.au/people/pulsar/psrcat} to identify  317 separate observations of 48 distinct pulsars.  Using the RRATalog\footnote{http://astro.phys.wvu.edu/rratalog}, we were also able to identify ten observations of nine distinct, known RRATs.

We combined these 327 known pulsar DPGs with a random sample of non-pulsar DPGs to create a fully labelled, benchmark data set of 10,000 total instances. We then used the benchmark data set to build and evaluate our machine learning classification models, as described in Section~\ref{sect:benchres}. Finally, we used the classification models with the best performance on the benchmark to classify every instance in the full data set (see Section~\ref{sect:fullres}).

\subsection{DPG Classification with Machine Learning}
\label{sect:learning}

This section describes the particular machine learning algorithms we used, how we dealt with the imbalance inherent to the data, and how we evaluated the performance of our classification models.

\subsubsection{Machine Learning Algorithms}
\label{sect:alg}

We used six machine learning algorithms of different types: an artificial neural network, support vector machine, direct rule learner, standard tree learner, hybrid rule-and-tree learner, and ensemble tree learner. The intent of choosing different types of learners was to see if any certain machine learning technique performs better overall when searching for pulsars in single-pulse search results. Each learner is listed in Table~\ref{tab:learners}. For this work, we used learners' implementations available through \textit{Weka}, a popular machine learning software suite \citep{weka}.

Artificial neural networks (ANNs) have been used in several related papers working with periodicity searches \citep{2010MNRAS.407.2443E,2012MNRAS.427.1052B,2014ApJ...781..117Z,2014MNRAS.443.1651M}, as mentioned in Section~\ref{sect:rw-periodicity}. The ANN we used is the Java implementation of a Multilayer Perceptron Network (MPN), which classifies instances using the supervised learning method of back-propagation and a sigmoid activation function in all neural nodes \citep{mpn}.

Support Vector Machines (SVMs) are a class of supervised learners which create higher order decision boundaries, called hyperplanes, to separate different instances by class. They use mapping functions, called kernels, to transform the input space into a more easily separable feature space. To construct an optimal hyperplane to separate the instances in this transformed space, SVMs use iterative training algorithms to minimize an error function. Sequential minimal optimization (SMO) is a Java implementation of a support vector machine \citep{Platt1998,Keerthi2001}. SMO solves the optimization problem of minimizing error by a divide and conquer strategy, breaking the problem into a series of smallest possible problems which are then solved analytically. 

The direct rule learner tested was JRip, the Java implementation of the RIPPER \citep{jrip}. As a rule learner, JRip creates a set of rules from the training set and then classifies each instance in the test set based on the generated rules. The rules consist of one or more antecedents followed by a single consequent, following a basic ``if \textit{antecedent}(s) then \textit{consequent}'' structure. Rule learners follow a ``separate and conquer'' methodology, i.e., they build a rule that covers as many instances as possible, remove all instances for which that rule is true from the training set, then continue this process recursively until all instances are covered by at least one rule.

The standard tree learner we tested was J48, the Java implementation of the C4.5 \citep{j48} learner. Decision tree algorithms approach classification with a ``divide-and-conquer'' strategy. They operate by determining what criteria \textit{best} divides the test set into separate groups. J48 uses a normalized function called \textit{information gain}, which is defined in terms of information content, or \textit{entropy} \citep{Russell:2003:AIM:773294}.

PART is a hybrid learner developed using ideas from both decision tree and rule learners \citep{part}. PART adopts the separate-and-conquer strategy of building sets of rules, but differs in the way individual rules are created. To make each rule, rather than incrementally adding antecedents one at a time, PART builds a pruned decision tree for the current set of instances and makes a rule from the leaf with the greatest coverage, discarding the rest. PART takes its name from this method of generating PARTial trees to create rules, and gains simplicity while saving time by removing the global optimization step.

Finally, we used an ensemble tree learner called RandomForest (RF) \citep{randomforest}. RandomForest uses an ensemble of decision trees to classify instances. For each tree, a random vector of attributes is selected from the training set and used to make the decisions at each node. In a RandomForest, each attribute vector in the set of random vectors is independent and identically distributed. To classify an unknown instance, the instance is inputted to each tree in the forest and each tree votes on the class of the instance. The instance is then assigned the class with the most votes. RandomForests are well suited for astronomical searches for their reported accuracy, efficiency in handling large data sets, and robustness with respect to noise.

\begin{table}
	\caption{The name and type of each machine learning algorithm used for this work.}
	\centering
	\footnotesize
		\begin{tabular}{||c|c|c||}
		\hline
		\hline
		\textbf{Learner} & \textbf{Type} \\
		\hline
			MPN & Artificial Neural Network \\
			SMO & Support Vector Machine \\
			JRip & Rule \\
			J48 & Tree \\
			PART & Rule + Tree \\
			RandomForest & Ensemble Tree \\
		\hline
		\hline
		\end{tabular}
		\label{tab:learners}
\end{table}

\subsubsection{Multiclass Classification}
\label{sect:mc}
\begin{figure*}
	\centering
	\begin{tabular}{c c}
		\textbf{Pulsar} & \textbf{Very Bright pulsar} \\
		\includegraphics[width=0.48\linewidth]{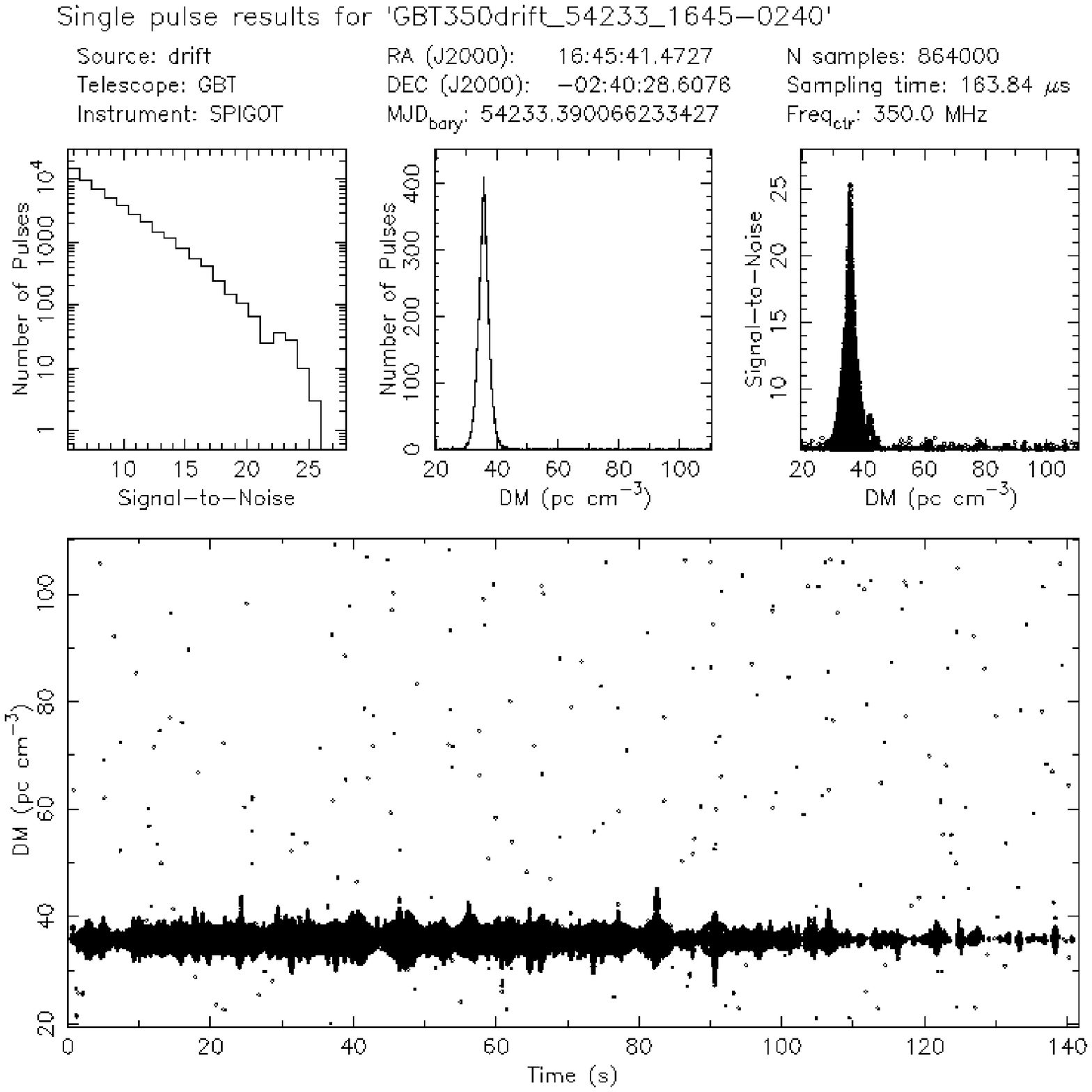} &	\includegraphics[width=0.48\linewidth]{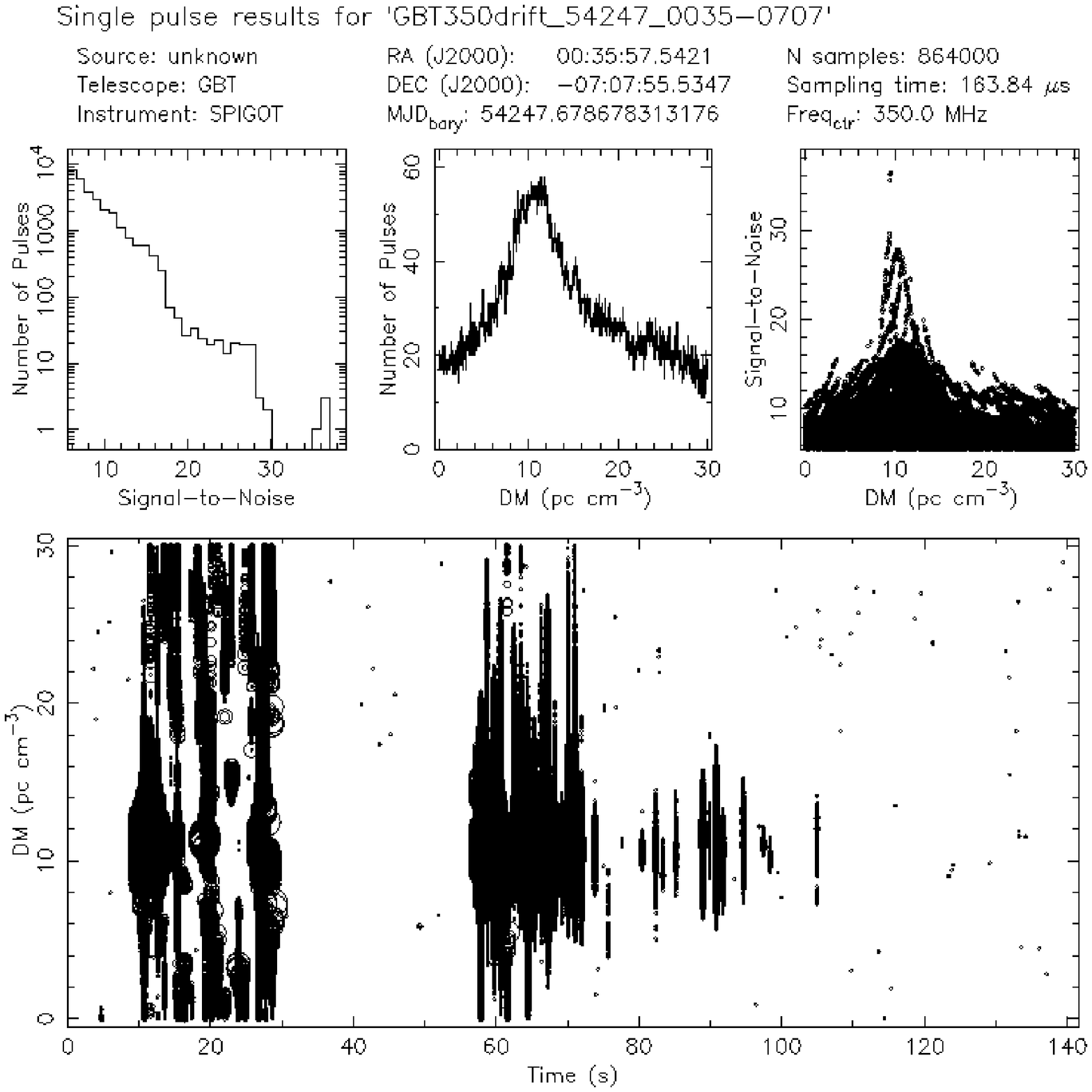} \\
		\textbf{RRAT-like pulsar or FRB} & \textbf{Non-pulsar} \\
		\includegraphics[width=0.48\linewidth]{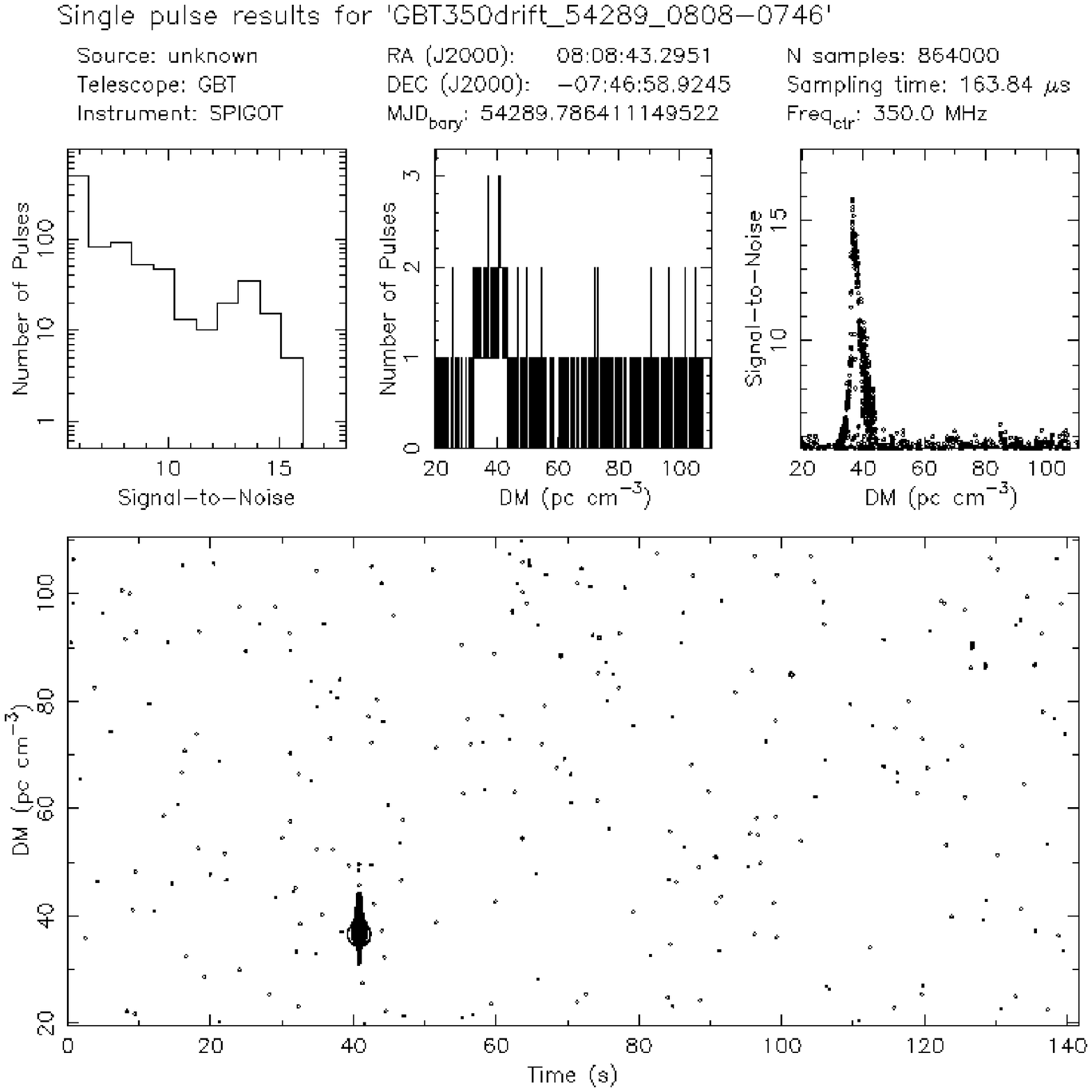} & \includegraphics[width=0.48\linewidth]{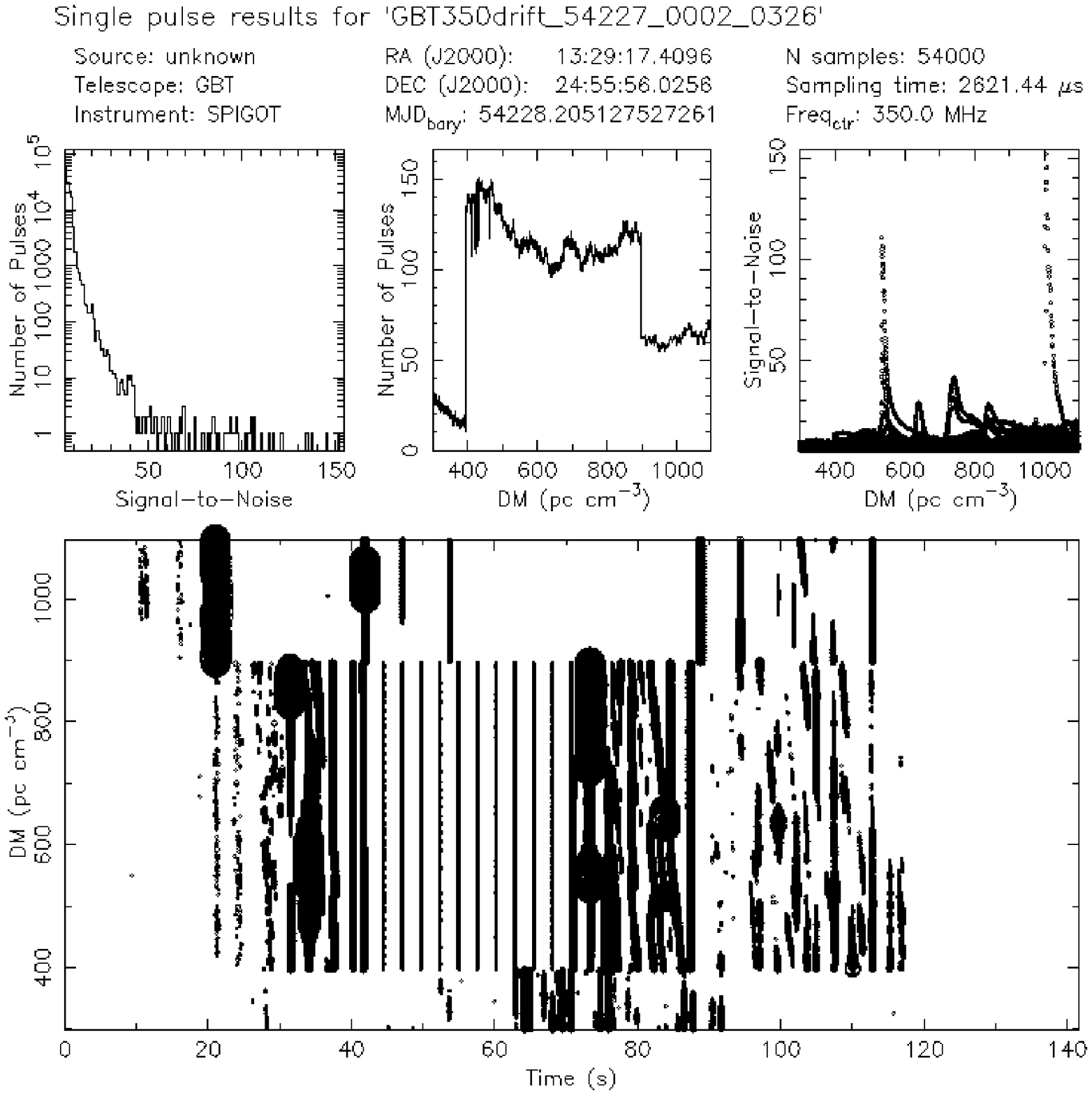} \\
	\end{tabular}
\caption{Four classes of pulsars based on the appearances of their candidate plots.}
\label{fig:mcplots}
\end{figure*}

Binary classification occurs when the class variable can assume one of two values, e.g. pulsars and non-pulsars. In multiclass classification, more specialized models can be created by training on multiple classes, each consisting of similar instances. In addition to binary classification models, we also used multiclass versions of the learners presented in Section~\ref{sect:alg}. To accomplish this, we divided our training examples into four classes based on their appearances: pulsars, very bright pulsars, RRAT-like pulsar or FRB, and non-pulsars. Each DPG can belong to a candidate plot of one of these four classes. Figure~\ref{fig:mcplots} provides examples of each class of candidate plot. Compared to pulsars, plots for very bright pulsars are often missing the brightest pulses at the DM of the pulsar, resulting in a flatter distribution at the peak of the SNR vs DM subplot. This is due to $single\_pulse\_search.py$ clipping the bright pulses. While RRAT-like pulsars have the same shape as pulsars in the SNR vs DM plot, their lack of sustained emission causes them to have lower values for certain metrics, such as \textit{IntegratedSNR}. FRBs appear similar to RRAT-like pulsars, with only one pulse at high DM.

\subsubsection{Imbalance Considerations}
\label{sect:balance}
In data classification, the majority of data gathered is often not very interesting, (e.g. regular usage in network security or financial transactions) or it is mostly interference or noise (e.g. RFI in pulsar searching).  When a data set has a very skewed distribution of class variables, it is said to be \textit{imbalanced}.  For our data, out of more than 1.5 million instances in the full data set, only 327 were positive examples of the target class. The benchmark data set with no imbalance treatment consisted of the 327 positive examples with 9,673 randomly selected and manually validated negative examples. With such a miniscule ratio of minority to majority class members, many learners will ``over-train'' on the majority class. Therefore, we also considered three versions of the benchmark with three different imbalance treatments: 
\begin{itemize}
	\item \textit{Oversampled} -- Random selections are made from the minority class (with replacement, i.e., the same example may be chosen multiple times) in order to improve the balance between the minority and majority class. 
%
	\item \textit{SMOTE} -- Synthetic Minority Oversampling TEchnique is similar to oversampling, but each time a random member of the minority class is selected after the first, a synthetic instance is created with small, random perturbations in the values of each of its features. This technique was designed to help eliminate the problem of overfitting a learner to the minority class members that are oversampled \citep{chawla2002smote}.
	\item \textit{Undersampled} -- A traditional treatment to the imbalance problem, where a random sample of the majority class is combined with all instances of the minority class \citep{sampling}. 
\end{itemize}

\subsubsection{Learning process}
\label{sect:learning-process}

We evaluated the performance of the six learners shown in Table~\ref{tab:learners} on the imbalanced benchmark data set described in Section~\ref{sect:labels} and on three additional benchmark data sets created using the imbalance treatments described in Section~\ref{sect:balance}. We use the term \textit{classifier} to refer to the combination of a machine learning algorithm trained on a specific benchmark data set. 

For the evaluation, we chose five fold cross-validation, which divides each benchmark version into five folds. The folds contain stratified random samples, i.e., the positive examples are divided equally among them. Four folds were used to train the learner (the ``training set'') and the fifth was used to test the learner's classifications (the ``test set''). Five trials were performed with a different fold serving as the testing set for each trial. 

When using oversampling imbalance treatments with cross-validation, precautions must be taken to maintain mutual exclusion between the training and testing sets. Otherwise, the same positive examples may exist in both the training and testing sets and the learners may falsely appear to perform very well in the testing environment because they are not being tested on unseen data. 
We avoided this by first dividing the data into folds and then applying the imbalance treatment only to the training set, and testing the learner on the fifth, unchanged fold which was held out as a testing set. 
The advantages of performing evaluations in this manner are that all observations are guaranteed to be used for both training and testing, learners are tested on unseen data, and each observation is used for testing exactly once.


\subsubsection{Metrics for Evaluation of the Classifications}
\label{sect:perf}

\begin{figure}
\centering
	\includegraphics[width=0.6\linewidth]{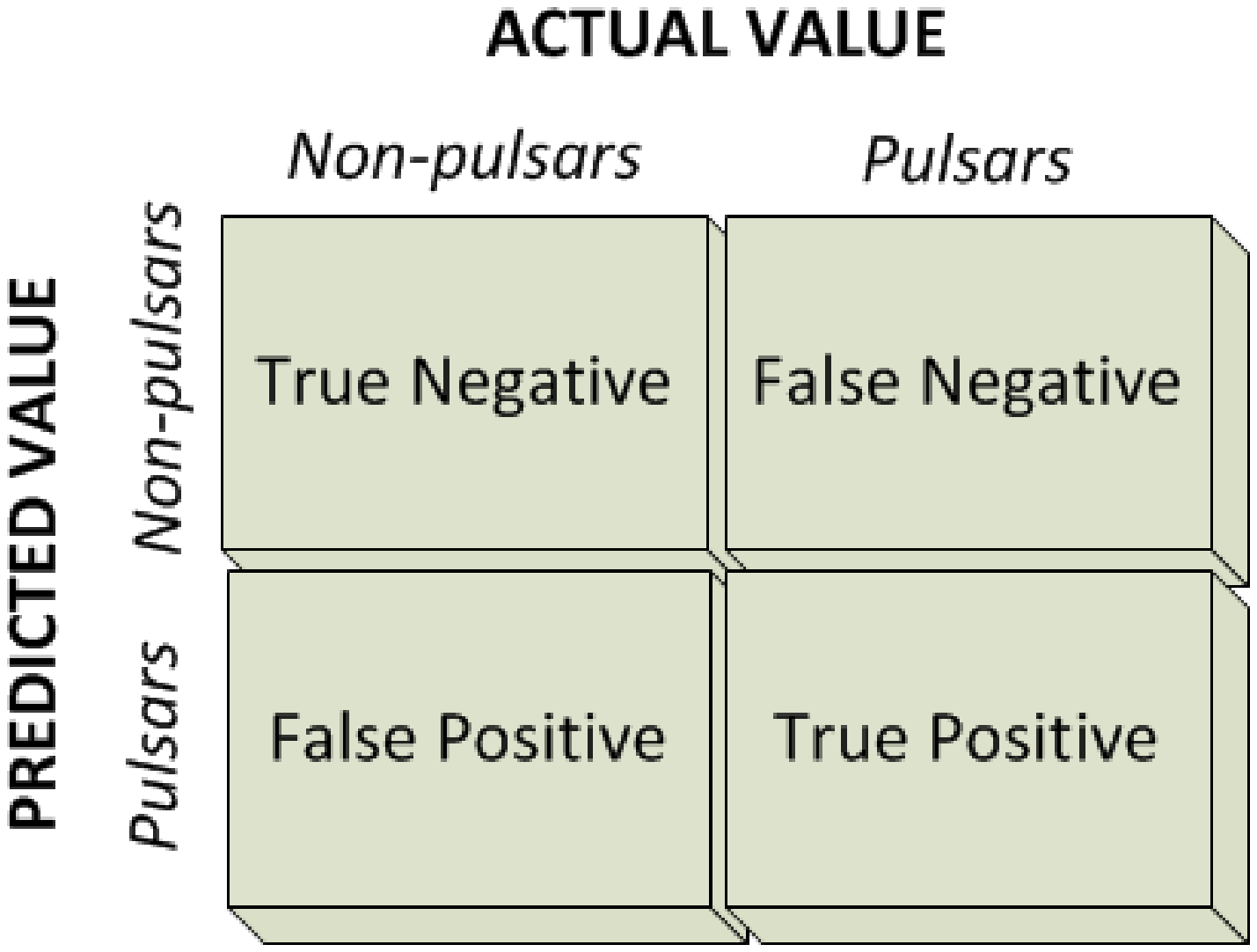}
\caption{Confusion matrix for pulsar classification. The rows represent the predicted class value of the model, and the columns represent the actual values. }
\label{fig:conf}
\end{figure}

To evaluate the effectiveness of our classifiers, we used several performance metrics calculated from \textit{confusion matrices} \citep{Witten:2005:DMP:1205860}. A confusion matrix is a summary table of a classifier's performance on a given test set. In the confusion matrix for binary classification shown in Figure~\ref{fig:conf}, the predicted values are represented by the rows $non$-$pulsars$ and $pulsars$. The actual values are represented by the columns $non$-$pulsars$ and $pulsars$. The result of any classifications then reside in one of the following four boxes\footnote{A confusion matrix can only be computed on a fully labelled data set. If unlabelled instances exist in a data set, they cannot be placed within the confusion matrix and other criteria must be used for evaluation.}:

\begin{itemize}
  \item \textit{True Negatives} (TN) -- represent the number of DPGs that were non-pulsars and were correctly classified as non-pulsars,
	\item \textit{False Negatives} (FN) -- represent the number of DPGs that were pulsars, but were incorrectly classified as non-pulsars,
	\item \textit{False Positives} (FP) -- represent the number of DPGs that were non-pulsars, but were incorrectly classified as pulsars, and
	\item \textit{True Positives} (TP) -- represent the number of DPGs that were pulsars and were correctly classified as pulsars.
\end{itemize}

For pulsar classification, we are most concerned with the number of true positives and false negatives. \textit{Recall} is a performance measure that quantifies the ability of the classifier to correctly classify the positive training instances:
\begin{equation}
Recall = \frac{TP}{TP + FN}.
\end{equation}
A perfect recall, or true positive rate (TPR) of 1, indicates that all positively labeled instances are properly classified.

The \textit{false negative rate} (FNR) is the complement of the TPR, and represents the conditional probability of mis-classifying real pulsars as non-pulsars, which is very undesirable. It is defined as:
\begin{equation}
FNR = \frac{FN}{FN+TP}.
\end{equation}

The \textit{precision} describes what fraction of the positive classifications are relevant, and is defined as:
\begin{equation}
Precision = \frac{TP}{TP + FP}.
\end{equation}
A perfect precision of 1 means that every instance predicted to be positive was actually  a positive instance.

The \textit{false positive rate} of a classifier describes how often the classifier `cried wolf', or falsely labelled a negative instance. It is defined as:
\begin{equation}
FPR = \frac{FP}{TN+FP}.
\end{equation}
A classifier with a high false positive rate will result in wasted effort to manually inspect non-pulsar DPGs, but is more desirable than a high FNR.

For our experiments, we also report the harmonic mean between the recall and precision, commonly known as the \textit{F-measure} (F-M), which is defined by:
\begin{equation}
\mbox{F-M} = 2*\frac{Precision*Recall}{Precision + Recall}.
\end{equation}
The F-measure has a high value if both the recall and precision are high. This makes it particularly suitable for evaluating the effectiveness of an automated pulsar classifier because it characterizes the ability of a classifier both to not miss pulsars, and to produce fewer false positives that require manual inspection. A perfect learner would have a value of 1 for its F-measure.
\section{Results}
\label{sect:results}

In our experiments, we assigned RAPID a bin size of 25 points and a slope threshold of 0.5. We chose these parameters for our initial study for the following reasons: (a) the bin size was large enough to smooth over noise, yet small enough to detect our DPG examples, (b) the slope threshold was shallow enough to catch the more gradual slopes of some of the wider DPG examples, and (c) our preliminary experimentation with these values identified most known pulsar signals in our data. We ran RAPID with these parameters over the 42,405 observations from the GBT drift-scan survey, which resulted in 1,578,789 DPGs. Since a DPG is any noticeable peak in the DM vs SNR subplot of a candidate plot, and there are many such peaks, there are significantly more DPGs than observations. We intentionally selected a bin size and slope threshold that resulted in a large number of DPGs in order to decrease the probability of missing any pulsars.

This section is divided into three subsections. First, we present the results from training and testing our learners on the four versions of the benchmark data set. Next, we present the results from classifying each DPG in the full data set using our best classifiers. Finally, we compare the results based on one of the best classifiers with results produced by a simple threshold ranking system.

This section uses the following notation to refer to a given classifier:
\begin{equation}
	[learner]_{[treatment]}^{[classes]},
\end{equation}
where $learner$ is an abbreviation for the machine learning algorithm, $treatment$ is the imbalance treatment used, and $classes$ is either `2' for binary classification or `4' for multiclass classification. For example, the notation $RF_{over}^2$ refers to the classification model created by training a binary class RandomForest machine learning algorithm on the benchmark data set with the oversampled imbalance treatment.

\subsection{Results Based on the Benchmark Data Sets}
\label{sect:benchres}

\begin{table*}
	\centering
	\footnotesize
	\caption{The benchmark results for our classifiers. The centre columns report mean values for the performance metrics described in Section~\ref{sect:perf}. The final two columns report the average time taken to train and test the learners.}
	\begin{tabular}{||l|r|r|r|r|r|r|r||}
		\hline
		\hline
		\textbf{Classifier} & \textbf{Recall} &  \textbf{FNR} &  \textbf{Precision} & \textbf{FPR} & \textbf{F-M} & \textbf{Train(s)} & \textbf{Test(s)} \\
		\hline
		$MPN_{none}^{2}$	&	0.238	&	0.762	&	0.654	&	0.004	&	0.349	&	16.818	&	0.008	\\
		$SMO_{none}^{2}$	&	0.000	&	1.000	&	0.000	&	0.000	&	0.000	&	3.324	&	0.002	\\
		$JRIP_{none}^{2}$	&	0.571	&	0.429	&	0.680	&	0.009	&	0.620	&	1.269	&	0.005	\\
		$PART_{none}^{2}$	&	0.548	&	0.452	&	0.723	&	0.007	&	0.624	&	0.353	&	0.001	\\
		$J48_{none}^{2}$	&	0.517	&	0.483	&	0.689	&	0.008	&	0.591	&	0.170	&	0.004	\\
		$RF_{none}^{2}$	&	0.459	&	0.541	&	0.918	&	0.001	&	0.612	&	2.901	&	0.053	\\
		\hline															
		$MPN_{over}^{2}$	&	0.867	&	0.133	&	0.199	&	0.135	&	0.324	&	16.301	&	0.017	\\
		$SMO_{over}^{2}$	&	0.602	&	0.398	&	0.120	&	0.172	&	0.200	&	0.545	&	0.016	\\
		$JRIP_{over}^{2}$	&	0.739	&	0.261	&	0.466	&	0.033	&	0.572	&	3.325	&	0.015	\\
		$PART_{over}^{2}$	&	0.706	&	0.294	&	0.462	&	0.032	&	0.558	&	0.814	&	0.015	\\
		$J48_{over}^{2}$	&	0.689	&	0.311	&	0.430	&	0.035	&	0.529	&	0.254	&	0.013	\\
		$RF_{over}^{2}$	&	0.718	&	0.282	&	0.714	&	0.011	&	0.716	&	2.931	&	0.091	\\
		\hline															
		$MPN_{smote}^{2}$	&	0.878	&	0.122	&	0.222	&	0.120	&	0.354	&	16.420	&	0.013	\\
		$SMO_{smote}^{2}$	&	0.749	&	0.251	&	0.104	&	0.251	&	0.182	&	0.199	&	0.012	\\
		$JRIP_{smote}^{2}$	&	0.852	&	0.148	&	0.362	&	0.058	&	0.509	&	3.075	&	0.009	\\
		$PART_{smote}^{2}$	&	0.842	&	0.158	&	0.344	&	0.062	&	0.488	&	1.147	&	0.017	\\
		$J48_{smote}^{2}$	&	0.823	&	0.177	&	0.351	&	0.059	&	0.492	&	0.428	&	0.011	\\
		$RF_{smote}^{2}$	&	0.834	&	0.166	&	0.538	&	0.028	&	0.654	&	5.503	&	0.114	\\
		\hline															
		$MPN_{under}^{2}$	&	0.884	&	0.116	&	0.162	&	0.173	&	0.274	&	1.529	&	0.024	\\
		$SMO_{under}^{2}$	&	0.786	&	0.214	&	0.087	&	0.319	&	0.157	&	0.019	&	0.017	\\
		$JRIP_{under}^{2}$	&	0.896	&	0.104	&	0.205	&	0.135	&	0.334	&	0.073	&	0.014	\\
		$PART_{under}^{2}$	&	0.895	&	0.105	&	0.171	&	0.168	&	0.288	&	0.027	&	0.011	\\
		$J48_{under}^{2}$	&	0.891	&	0.109	&	0.198	&	0.140	&	0.324	&	0.241	&	0.079	\\
		$RF_{under}^{2}$	&	0.927	&	0.073	&	0.287	&	0.090	&	0.438	&	0.241	&	0.079	\\
		\hline
		\hline
	\end{tabular}
	\label{tab:results-all}
\end{table*}

\begin{figure*}
	\centering
	\includegraphics[width=1.0\linewidth]{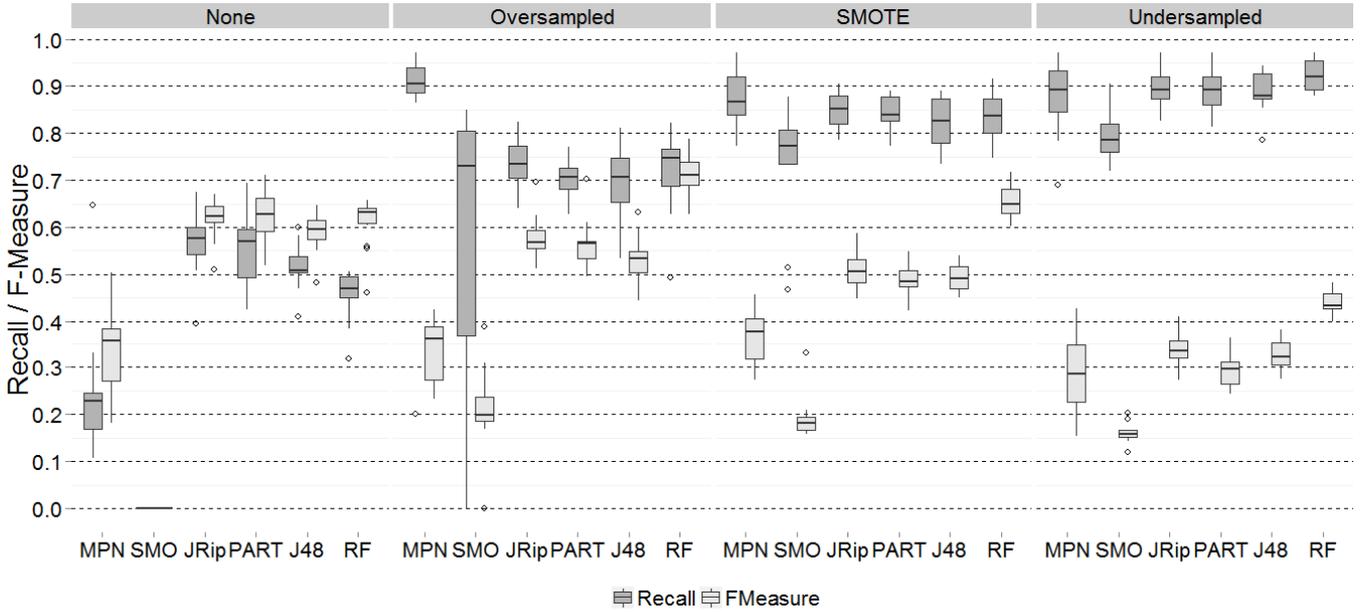}
	\caption{The results of binary machine learning trials on the four versions of the benchmark data set. The median of each distribution is the central horizontal line, the shaded areas to either side illustrate the semi-interquartile ranges, the lines or ``whiskers'' extending from each end give the first and fourth quartiles, and outliers are represented by open circles. The distributions of performance metrics (shown on the y axis) for each learner (annotated at the bottom) are organized into four groups, one for each version of the benchmark data set (annotated at the top).}
	\label{fig:boxplots}
\end{figure*}

We used the six learners shown in Table~\ref{tab:learners} to build classifiers for each of the four versions of the benchmark data set in three repeated trials using five fold cross validation for a total of 360 trials. The trials were conducted using Weka's experimenter on an Alienware M14xR2 with an Intel Core i7 CPU, 16 GB RAM, and a 512 GB solid state drive. The results of the binary classification are displayed in Table~\ref{tab:results-all}, which includes training and testing times. Figure~\ref{fig:boxplots} shows boxplots of the distributions of key performance metrics, grouped by the benchmark version.

Based on Table~\ref{tab:results-all} and Figure~\ref{fig:boxplots}, we make the following observations:

\begin{itemize}
	\item All classifiers with imbalance treatments had higher recall values than those with unbalanced data.
	\item RandomForest provided the highest average F-measure and good recall values on all data sets with imbalance treatments.
	\item MPN had the best recall values for the oversampled and SMOTE imbalance treatments, but the second worst F-measure for all imbalance treatments.
	\item SMO had the worst performance for the four benchmark data sets.
\end{itemize}

The choice of a best classifier from the benchmark trials depends on the most desirable performance measure. For automating pulsar classification, the F-measure may be considered the most important performance measure because a classifier with a high F-measure must have good scores for both recall and precision. (As described in Section~\ref{sect:perf}, a high recall indicates the classifier will correctly classify most positive instances and a high precision indicates the classifier will not result in many false positives.) While a high recall is important, the point of automation is to minimize human involvement. A low precision means that only a small fraction of positive classifications is relevant, that is, there are many false positives which would require manual inspection and therefore is undesirable for this work. 
For example, although the $MPN_{over}^{2}$ classifier had the best recall among learners on the oversampled benchmark, it has a very low F-measure because it produced many false positives. 
Therefore, with respect to the F-measure the best classifiers are $RF_{smote}^{2}$ and $RF_{over}^{2}$.


It should be noted that the training and testing times show that MPNs are by far the slowest of the six learners tested. MPNs, as most ANNs, use a gradient descent optimization routine to determine the weighted values between network nodes during back propagation. Gradient descent calculations are computationally expensive, and are often the cause of increased training times. Furthermore, while recall values were very high, the F-measures obtained for MPNs were consistently lower than all learners except SMO, with a large variance. Note that ANNs are one of the most common machine learning techniques applied to the problem of radio pulsar detection in periodicity searches, and were used in each paper discussed in Section~\ref{sect:rw-periodicity} that performed machine learning \citep{2010MNRAS.407.2443E,2012MNRAS.427.1052B,2014ApJ...781..117Z,2014MNRAS.443.1651M}.

Results from building and testing multiclass learners on the four versions of the benchmark data set were similar to the binary classification results, with the $RF_{smote}^{4}$ and $RF_{over}^{4}$ classifiers performing the best with respect to F-measure. 

\subsection{Results Based on the Full Data Set}
\label{sect:fullres}

Based on the results reported in Section~\ref{sect:benchres} we selected the models produced by two learners -- RF (best F-measures) and MPN (best recalls) -- in combination with all imbalance treatments to classify every instance in the full data set. Since most of the DPGs in the full data set were not labelled, it was not possible to calculate the same performance metrics as for the benchmark data sets. Instead, we evaluated the performance of the models by the following criteria: how many potential discoveries (PDs) were found, how many known pulsars were classified correctly (CKs), how many additional known pulsars (AKs) were found beyond those included in the benchmark, and how many DPGs classified as pulsars were false positives (FPs), i.e., non-pulsars incorrectly classified as pulsars \footnote{Note that not all instances in the FPs column were examined for classifiers with more than 9,000 FPs. Such classifiers were able to achieve high PDs, CKs, and AKs by simply classifying almost everything as a pulsar, which defeats the purpose of automation.}. Table~\ref{tab:fullres} provides the results for all benchmark versions of the binary RF and MPN learners.


\begin{table}
	\centering
	\footnotesize
	\caption{A comparison of the performance of classifiers using binary RandomForest (RF) and Multilayer Perceptron Network (MPN) learners on the full data set.}
		\begin{tabular}{||l|r|r|r|r||}
			\hline
			\hline
			\textbf{Classifier} & \textbf{PDs} & \textbf{CKs} & \textbf{AKs} & \textbf{FPs} \\
			\hline
				$RF_{none}^{2}$ & 0 & 304 & 5 & 32 \\
				$RF_{over}^{2}$ & 2 & 327 & 15 & 451 \\
				$RF_{smote}^{2}$ & 3 & 326 & 46 & 1,940 \\
				$RF_{under}^{2}$ & 6 & 326 & 33 & 9,750 \\
			\hline
				$MPN_{none}^{2}$ & 0 & 79 & 1 & 696 \\
				$MPN_{over}^{2}$ & 6 & 309 & 23 & 43,943 \\
				$MPN_{smote}^{2}$ & 3 & 257 & 23 & 14,066 \\
				$MPN_{under}^{2}$ & 6 & 298 & 29 & 110,629 \\
			\hline
			\hline
		\end{tabular}
	\label{tab:fullres}
\end{table}

Three important results stand out from Table~\ref{tab:fullres}: 
(1) RF models had almost perfect CK and 1-2 orders of magnitude fewer FPs than their MPN counterparts. 
This finding was expected based on the low F-measures of the MPN learner on the benchmark data sets. 
(2) Classifiers using MPN learners had lower CK, i.e., they failed to correctly classify from $5-75\%$ of the known pulsar examples. This result was unexpected, as the classifiers with MPN learners had the highest recall values in the benchmark experiments. 
(3) The most FPs, for both RF and MPN, were produced in combination with the undersampled imbalance treatment, which is consistent with the lowest precision and F-measure obtained on the benchmark data set. 

\begin{figure}
	\centering
	\includegraphics[width=1.0\linewidth]{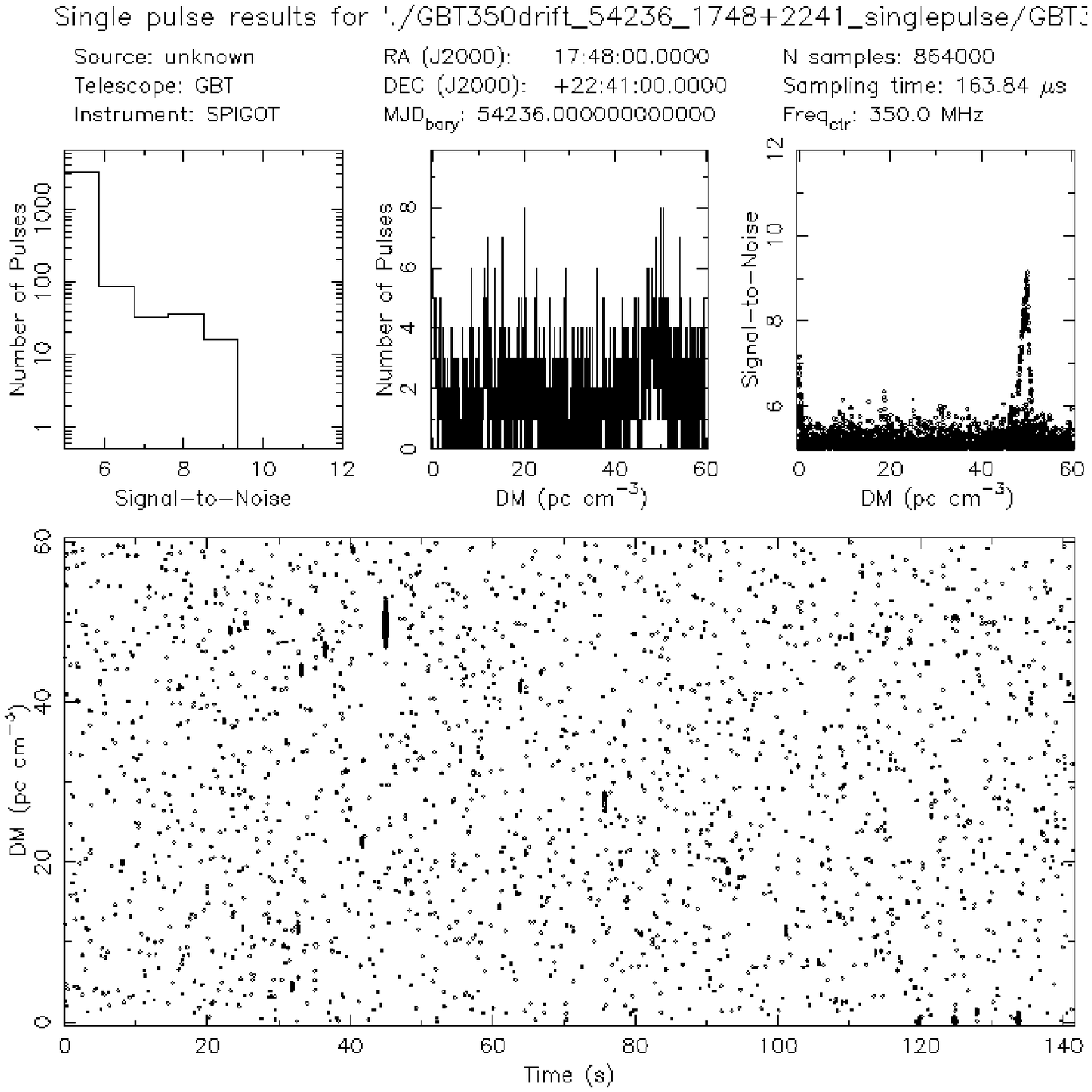}
	\caption{One potential RRAT discovery found by our classifiers.}
	\label{fig:newrrat}
\end{figure}

Based on the results presented in Table~\ref{tab:fullres}, we decided to use the $RF^4_{over}$ and $RF^4_{smote}$ multiclass classifiers on the full data set, due to their nearly perfect CK, high AK, and low FP values. 
($RF^4_{under}$ and $MPN$ classifiers with all imbalance treatments were not used due to the high number of FPs.)
Table~\ref{tab:fullresmc} reports the results, from which we make the following observations: 
(1) The classifiers with multiclass RF learners were superior to their binary counterparts, for both imbalance treatments, because they were trained on three pulsar classes whose appearance and feature values are quite different. Specifically, classifiers using multiclass RF learners were able to detect potential RRATs, like the one shown in Figure~\ref{fig:newrrat}, which were missed by the binary classifiers. 
(2) $RF^4_{smote}$ found six PDs, which contained all of the PDs from the other classifiers, with a much smaller number of FPs than the binary classifiers shown in Table~\ref{tab:fullres}.  
(3) With respect to imbalance treatments, there is a tradeoff to be made. Compared to the oversampled treatment, SMOTE resulted in more detections, both PDs and AKs, but with the added cost of over five times more FPs requiring manual inspection.


\begin{table}
	\centering
	\footnotesize
	\caption{A comparison of the performance of oversampled and SMOTE multiclass RandomForest (RF) classifiers on the full data set.}
		\begin{tabular}{||l|r|r|r|r||}
			\hline
			\hline
			\textbf{Classifier} & \textbf{PDs} & \textbf{CKs} & \textbf{AKs} & \textbf{FPs} \\
			\hline
				$RF_{over}^{4}$ & 5 & 327 & 32 & 330 \\
			    $RF_{smote}^{4}$ & 6 & 316 & 35 & 1,718 \\
			\hline
			\hline
		\end{tabular}
	\label{tab:fullresmc}
\end{table}

\subsection{Comparison of Our Results to Simple Ranking}
\label{sect:evalres}

To further evaluate the performance of one of our classifiers, $RF_{over}^{2}$, we compared our results to a simple ranking approach based on the feature \textit{$SNRMax\chi^2$}, a measure of how well the shape of the DPG in the SNR vs DM plot fits the ideal, theoretical shape of a dispersed pulse given by Equation~\ref{eq:fit} \citep{0004-637X-596-2-1142}.

We sorted all DPGs in the full data set by their \textit{$SNRMax\chi^2$} values and calculated summary statistics ($1^{st}$ quartile, median (i.e., the $2^{nd}$ quartile), $3^{rd}$ quartile, mean, and maximum) to use as thresholds. We then examined how many DPGs in the full data set had \textit{$SNRMax\chi^2$} values less than each statistic, and how many of those DPGs were known pulsars. Note that no ranking system based on \textit{$SNRMax\chi^2$} can detect 100\% of the known pulsars in the data set, since the fitting routines for several known pulsars failed to converge due to noise spikes which offset their central peaks. 

We report the percentage of known pulsars detected with values below a given threshold, which we call the Known Detection Rate (KDR). To quantify how much effort would produce results if one performed manual inspection of all top ranked DPGs, we also computed what percent of the top ranking DPGs are known pulsars. We call this the Positive Effort Rate (PER). The results are presented in Table~\ref{tab:threshold}.

\begin{table}
 \caption{Rankings based on a simple threshold for the feature \textit{$SNRMax\chi^2$}. The first column gives the statistic used as the threshold value, the second column shows the value of \textit{$SNRMax\chi^2$} which will be used as the threshold, the third column displays the number of DPGs in the full data set that have a value below the threshold, the fourth column shows how many of the DPGs in the third column are known pulsars, the fifth column gives the percent of known pulsars detected below the given threshold, and the sixth column shows what percentage of the top ranked DPGs are pulsars (PER). For many DPGs, the fitting routine could not reach convergence.}
	\centering
	\footnotesize
		\begin{tabular}{||l|r|r|r|r|r||}
			\hline
			\hline
			\textbf{Statistic} & \textbf{Value} & \textbf{\#<Value} & \textbf{\#DPGs} & \textbf{KDR} & \textbf{PER} \\
			\hline
			\textit{1st Quartile}	& 47 & 80,512 & 136  & 42\% & 0.17\% \\
			\textit{Median}	& 189 & 161,854 & 266  & 81\% & 0.16\% \\
			\textit{3rd Quartile}	& 571 & 242,534 & 312 & 95\% & 0.13\% \\
			\textit{Mean}	& 816 & 265,738 & 316  & 97\% & 0.12\% \\
			\textit{Maximum}	& $9\times{10^6}$ & 323,447 & 323  & 99\% & 0.10\% \\
			\hline
			\hline
		\end{tabular}
 \label{tab:threshold}
\end{table}

As Table~\ref{tab:threshold} shows, if the median value of \textit{$SNRMax\chi^2$} was used as a threshold, we would have to manually inspect over 160,000 DPGs. Only 0.16\% of those 160,000 DPGs would be known pulsars and we would only be able to detect 81\% of the total known pulsars in the data set. In comparison, if we used our binary oversampled RandomForest model, 100\% of the known pulsars will be correctly classified and less than 470 DPGs would require manual inspection. The final column in Table~\ref{tab:threshold} shows that with any threshold value, at best, less than $0.2\%$ of the top ranked DPGs will be known pulsars. Alternatively, our binary oversampled RandomForest model resulted in a PER of $41\%$. We believe that our machine learning approach outperforms the ranking because the classification models are multivariate, i.e., they take many different features of the DPGs into consideration.
\section{Conclusions}
\label{sect:conc}

In this paper, we presented the first machine learning approach to pulsar classification in single-pulse searches. The approach consists of two main stages: DPG identification and DPG classification. We used a novel peak identification algorithm, RAPID, to successfully identify DPGs, which are local peaks in the output from single-pulse searches, and extracted meaningful features to describe them. Then, we used machine learning algorithms with imbalance consideration to classify the identified DPGs, first on a benchmark data set and then on the full, unlabelled data set created based on observations made by the Green Bank Telescope. The benchmark data set was created with over three hundred known pulsar signals and over 9,600 manually validated negative examples. To examine the problem of imbalance, we applied three different imbalance treatments to the original unbalanced benchmark data set. We used these four versions of the benchmark (the original unbalanced version and the three balanced versions) to train and test binary and multiclass versions of six different machine learning algorithms, resulting in 48 classifiers. We found that every classifier using an imbalance treatment provided higher recall values than the classifiers using unbalanced data. 
The classifiers using the RF ensemble tree learner provided the best overall balance between recall and precision (i.e., the highest F-measure values). On the other hand, the classifiers we tested using MPNs resulted in the highest recalls, but second worse F-measures and the longest training and testing times.

Based on these results we selected a subset of classifiers to search for potential pulsar discoveries in the full, unlabelled data set. The results showed that the multiclass RF classifiers significantly outperformed the binary classifiers. Specifically, they reported as many potential discoveries, were better in detecting potential RRAT discoveries, and produced less false positives than the binary classifiers. The oversampled and SMOTE imbalance treatments each had advantages and disadvantages. While the oversampled classifiers perfectly classified all known pulsar examples with very few false positives, they missed potential discoveries that were found by the SMOTE classifiers. The SMOTE classifiers, however, misclassified several known pulsar examples and produced four to five times more false positives. 
%
%
%
Overall, the combination of the multiclass RF learner with the SMOTE imbalance treatment was the most efficient -- it detected six potential pulsar discoveries with less false positives than any other classifier which also detected all six potential discoveries. The potential discoveries are currently under further review. Confirming them will require making frequency-time plots of the raw search data to confirm the broadband nature of any pulses and the expected $\nu^{-2}$ dependence of the dispersive delay, and then performing re-observations of these sky positions.


In future work, we plan to incorporate data from single pulses in DM vs time plots, like the one in Figure~\ref{fig:plot1}, into our approach with the goal of improving its sensitivity to fainter pulses or pulses that may be obscured RFI. We also plan to explore additional aspects of multiclass learning.

While expert knowledge and manual inspection will always play a strong role in the pulsar search process, semi-automated machine learning approaches, such as the one presented in this paper, have great potential for future discoveries in radio astronomy. As radio telescopes become bigger and better, they will gather more data faster. This increase in data volume will make manual inspection of every candidate impossible. Intelligent, scalable search techniques are the only viable solution to the big data problems looming on the horizon for radio astronomers.

\section{Acknowledgements}
\label{sect:ack}
The work was supported by the NSF awards \#1327526 and \#1458952. In addition, Thomas Devine's work was supported by the WVU Foundation's Ruby Distinguished Doctoral Fellowship. He would like to thank the WVU Foundation for their continued support and Dr. Jayanth Chennamangalam for his patient assistance.  We would also like to thank the Scientific Editor, Assistant Editor, and Reviewer for their helpful comments, and Weiwei Zhu for useful discussions.

\bibliographystyle{mnras}
\bibliography{apj-jour,TD_PhD_Dissertation}

\end{document}